\documentclass{article}
\usepackage{a4wide,amssymb,amsmath,cite}
\title{Variational Analysis of Deconfinement in Compact U(1) Gauge Theory}
\author{B. M. Gripaios\thanks{b.gripaios1@physics.ox.ac.uk}
\\\emph{Department of Physics - Theoretical Physics, University of Oxford,} 
\\ \emph{1, Keble Road, Oxford. OX1 3NP  UK
}
\\
\\ OUTP-02 41P 
}
\date{\today}
\begin{document}
\bibliographystyle{h-elsevier2}
\maketitle
\begin{abstract}
A variational method is used to analyse compact $U(1)$ gauge theory in $2+1$-dimensions 
at finite temperature, $T$, weak coupling, $g$ and where the fundamental magnetic monopoles have magnetic charge $2\pi n/g$.
The theory undergoes a critical transition from a confining phase at temperatures below $T_c=2g^2/n^2\pi$ to a 
deconfined phase at temperatures above $T_c$. The free energy and all its derivatives are continuous at $T_c$, indicative of the BKT
phase transition. 
The relevant gauge-invariant correlation functions decay exponentially at large distances. 
The spatial Wilson loop obeys the area law at all finite temperatures, even for the non-compact theory.
The case $n=2$ corresponds to the compact $U(1)$ theory considered as a low energy effective theory
for the spontaneously broken Georgi--Glashow model. 
The results in this case agree with those derived previously for compact $U(1)$ in this model
using dimensional reduction of the Lagrangian.


\end{abstract}

\section{Introduction}
A recent paper \cite{Kogan:2002yr} uses a variational method to study the deconfinement transition in QCD 
at finite temperature. The method mimics the Rayleigh-Ritz variational method in the Schr\"{o}dinger formulation of
quantum mechanics. There the standard procedure is to form an \emph{ansatz} for the ground state wavefunction, 
parameterized by some
free parameters, and to minimise the expectation value of the Hamiltonian with respect to the parameters. 
This provides an upper bound for the ground state (vacuum) energy. The method at finite temperature is analogous: 
in the canonical ensemble formulation of quantum statistical mechanics, one forms an \emph{ansatz} for the density matrix,
with free parameters, and minimises the expectation value of the Helmholtz free energy. This provides an upper bound
for the free energy at a given temperature.

This paper carries on the theme of \cite{Kogan:2002yr}, 
extending the variational analysis to the compact $U(1)$ gauge theory in $2+1$-dimensions at finite temperature.
This theory was shown to exhibit confinement at zero temperature, even at weak coupling, 
by Polyakov many years ago \cite{Polyakov:1977fu}. Essentially, the confinement is due to gauge field vortex defects 
which act in the partition function as a gas of interacting magnetic monopoles, causing Debye screening of the gauge field, 
a non-perturbatively small mass gap for the photon, an area law for the Wilson loop
and confinement of electric charges.
This was confirmed by a variational analysis \cite{Kogan:1995vb, Kovner:1998eg}. 
More recently, the theory has been studied at finite temperature \cite{Svetitsky:1982gs, Agasian:1998wv}
as the low energy limit of the Georgi--Glashow model ($SU(2)$ gauge field coupled to adjoint Higgs). 
At weak coupling $g$ and low temperature $T$, the theory can be understood as a two dimensional Coulomb gas,
undergoing the BKT phase transition \cite{Berezinskii:1971,Kosterlitz:1973xp} at temperature $T_c=\frac{g^2}{2\pi}$.\footnote{A more
recent work \cite{Dunne:2000vp} including the dynamics of $W^{\pm}$ Bosons in the G--G model has shown
that there is a phase transition at an even lower temperature corresponding to the unbinding of $W^{\pm}$.
This does not concern us here: we are interested only in the compact $U(1)$ aspect.} 
However, because the interaction strength
in the Coulomb gas picture is temperature dependent, the phase structure is \emph{reversed}, with a dilute plasma
of unbound monopoles below $T_c$ and a gas of bound magnetic dipoles above. This further implies the deconfinement of 
electric charges above $T_c$. As we shall see below, this phase transition is indeed apparent in the variational analysis
at finite temperature: for the compact $U(1)$ theory considered on its own, with fundamental magnetic monopoles\footnote{By
fundamental we mean that the smallest magnetic charge is $2\pi n/g$; integer multiples of this are of course allowed.}
of magnetic charge $2\pi n/g$, there is a phase transition at temperature $T_c = 2g^2/n^2\pi$ (eq'n \ref{T_c}) from a
confined phase (with finite photon mass gap) to a deconfined phase (zero mass gap). The mass gap vanishes continuously,
indicative of a critical transition. Furthermore, one can show that the free energy and \emph{all} its derivatives
are continuous at the phase transition (\ref{cont}). 
This is strongly indicative of the BKT phase transition. For the compact $U(1)$
theory considered as the low energy limit of the spontaneously broken Georgi--Glashow model, in which the 't Hooft--Polyakov
monopoles have fundamental charge $4\pi/g$ \cite{Kovner:1992fm}, the phase transition is at $T_c=\frac{g^2}{2\pi}$, as expected.

The analysis proceeds as follows. Since a variational method is only as good as its \emph{ansatz},
it is prudent to begin by studying the form of the density matrix in simpler theories, 
starting with the simplest of all, the harmonic oscillator. It is trivial to construct the density matrix in the
canonical ensemble in the energy basis, less so in the co-ordinate basis. 
But this is precisely the basis one needs for applying the variational method to theories where the spectrum is unknown. 
In \S\ref{sec:sho}, it is shown that the density matrix for the harmonic oscillator takes the form of a Gaussian.
This generalises to free scalar field theories upon replacing co-ordinates by fields.

Section \ref{sec:var} begins by outlining the details of the variational method. 
The variational \emph{ansatz} is taken to be a Gaussian, with arbitrary kernels. 
The minimisation of the free energy is then discussed; it is shown that the minimisation for the harmonic oscillator
picks out the density matrix constructed previously in \S\ref{sec:sho}. 

Section \ref{sec:cQED} discusses compact $U(1)$, dealing first
with the requirement of gauge invariance.
A straightforward modification of the \emph{ansatz} for the density matrix (following \cite{Kogan:2002yr}) projects
onto the gauge-invariant sector of states in the theory. 
Thus armed, we proceed to the variational analysis of compact $U(1)$, where things are complicated by the singular gauge defects
acting as magnetic monopoles. However, at weak coupling, the monopole fugacity (and photon mass)
are non-perturbatively small and one expects
that the density matrix will not be significantly changed from the non-compact case. This is indeed what happens: upon minimisation
of the free energy one finds a solution for the kernels (\ref{kers}) with a photon mass-squared
proportional to the (non-perturbatively small) monopole fugacity (\ref{mass}). 
It is further shown that (as the coupling tends to zero) the
variational method reproduces the expected behaviour of 
\emph{non-compact} $U(1)$, which is after all
just a gas of free photons with a single massless transverse degree of freedom (in $2+1$ dimensions).

Once the variational solution for the kernels has been obtained, 
it is straightforward to analyse the phase structure of the theory (\S\ref{sec:phase}).
Substitution of the kernels into the definition (\ref{fug}) of the monopole fugacity $z$ yields an implicit equation for
the fugacity (\ref{fugfug}). This equation has two solutions: the first is the trivial one $z=0$; 
the second has $z$ non-perturbatively small for temperatures below $T_c$, zero at $T_c$ and non-perturbatively large above $T_c$.
The free energy is minimised by the latter solution below $T_c$ and by the former solution above $T_c$. Finally, it is
shown that all derivatives of the fugacity (and hence the free energy) are continuous at $T_c$, a signature of the BKT phase transition.

In section \ref{sec:corrs}, the compact gauge invariant correlation functions and the spatial Wilson loop are calculated.
The Wilson loop has the area law at all non-zero temperatures, even for the non-compact theory.
This is confirmed in the Lagrangian approach to finite-temperature field theory.
\section{The harmonic oscillator revisited} \label{sec:sho}
Construction of the density matrix for the harmonic oscillator in the energy basis is trivial. 
Let the normalised energy eigenstates 
of the Hamiltonian $\mathcal{H}$
be $| n \rangle$ with energy $E_n = (n + \frac{1}{2}) \omega$ (henceforth, $\hbar = m = k_B = 1$). 
In the canonical ensemble, the density matrix (normalised to the partition function) is then
\begin{gather}
\rho = e^{-\mathcal{H}/T} = \Sigma_n e^{- E_n/T} |n \rangle \langle n |.
\end{gather}
In the co-ordinate basis, this becomes
\begin{gather} \label{shodens}
\rho (x,x') = \Sigma_n e^{-E_n/T} \langle x |n\rangle \langle n | x' \rangle.
\end{gather}
The Schr\"{o}dinger equation is
\begin{gather}
\left[-\frac{1}{2}  \frac{\partial^2}{\partial x^2}  + \frac{1}{2} \omega^2 x^2 - \left( n + \frac{1}{2} \right) \omega \right]
\langle x | n\rangle = 0,
\end{gather}
with solution
\begin{gather}
\langle x | n\rangle = \left( \frac{\omega}{\pi (2^n n!)^2} \right)^{\frac{1}{4}} 
H_n (\omega^{\frac{1}{2}} x ) e^{-\frac{\omega x^2}{2}},
\end{gather}
where $H_n$ is a Hermite polynomial.
Substituting in (\ref{shodens}) and using the formula \cite{Eboli:1988fm}
\begin{gather}
\Sigma_n \frac{1}{n!}
\left( \frac{z}{2} \right)^n H_n (x) H_n (y) = \frac{1}{\sqrt{1-z^2}} \exp{-\frac{z}{1-z^2} \left[(x^2 +y^2)z -2xy \right] },
\end{gather}
one obtains
\begin{gather} \label{shodens2}
\rho (x,x^{'}) = \left( \frac{\omega}{\pi} \right)^{\frac{1}{2}} \frac{e^{ -\omega / 2 T}}{\sqrt{1-e^{-2\omega / T}}}
\exp{-\frac{\omega}{2(1-e^{-2\omega / T})} \left[ (1+e^{-2\omega / T})(x^2 +x^{'2})  - 4 e^{-\omega / T} x x^{'} \right] }.
\end{gather}
As a check, taking the trace one obtains
\begin{gather} \label{partit}
\mathrm{tr} \rho = \int dx \; \rho(x,x) = \frac{e^{-\omega / 2 T}}{1 - e^{-\omega / T}},
\end{gather}
the standard partition function as required. It is clear that this result generalises to
free scalar fields, which consist of many such modes labelled by momentum $p$ with dispersion relation $\omega(p)$. 
The density matrix can be written as a product over all the modes, or equivalently as a sum over modes in the exponent.
Swapping co-ordinates $x$, $x^{'}$ for scalar fields $A$, $A^{'}$ and dropping the normalisation factor, we can write the
density matrix for scalar fields as
\begin{gather}
\rho [A,A^{'}] = \exp{-\frac{1}{2} \int \frac {dp}{2\pi} \;
\frac{\omega}{(1-e^{-2 \omega / T})} \left[ (1+e^{-2 \omega / T})(A^2 +A^{'2})  - 4  e^{- \omega / T} A A^{'} \right] }.
\end{gather}

We note two points at this stage. Firstly, at zero temperature,
the density matrix reduces to  
\begin{gather}
\exp{-\frac{1}{2} \int \frac {dp}{2\pi} \; \omega (A^2 +A^{'2})},
\end{gather}
in which terms coupling $A$ to $A'$ vanish. But this must correspond to a pure state: the ground state. So, as a general rule,
terms coupling  $A$ to $A'$ correspond to finite temperature thermal disordering effects.

Secondly, we see that the density matrix for free theories (\emph{i.e.} the harmonic oscillator) takes a Gaussian form.\footnote{This
is fortuitous, since then the ensemble average of observable $O$, 
which is $\mathrm{tr} \rho O / \mathrm{tr} \rho$, is a tractable Gaussian functional integral.}
Moreover, it is possible to establish the converse: all Gaussian density matrices correspond to free theories. To do this
write the most general Gaussian density matrix (which must be Hermitian and therefore symmetric in the real case) as
\begin{gather} \label{shoans}
\rho(x,x^{'}) = \exp{-\frac{1}{2} [x G^{-1} x + x^{'} G^{-1} x^{'} - 2 x H x^{'}]}.
\end{gather}
Then reverse the arguments of equations (\ref{shodens}-\ref{shodens2}), substituting $(G,H)$ for $(\omega,T)$ via
\begin{align} \label{shokers}
G^{-1} &= \omega \frac{1+e^{-2\omega / T}}{1-e^{-2\omega / T}} \nonumber \\
H &= 2 \omega \frac{e^{-\omega / T}}{1-e^{-2\omega / T}}.
\end{align}
It is thus evident that the most general Gaussian density matrix (\ref{shoans}) can be formally written as the operator
\begin{gather}
\rho = e^{-\mathcal{H}/T},
\end{gather}
where $\mathcal{H}$ is the Hamiltonian for an harmonic oscillator, with $(\omega,T)$ some functions of $(G,H)$ obtained by inverting
(\ref{shokers}). Note that this argument holds independent of thermal equilibrium considerations: $(\omega,T)$ are to be regarded
merely as a re-parameterisation (actually a diagonalisation) of $(G,H)$. 
This one-to-one identification between Gaussian density matrices and harmonic oscillators will turn out to be crucial in determining
the entropy corresponding to (\ref{shoans}). 
\section{The variational approach} \label{sec:var}
In the preceeding section, the density matrix has been derived for the harmonic oscillator and for the generalisation to
 free scalar fields. 
Now, for the purposes of pedagogy, let us derive the same results using the variational method. A suitable \emph{ansatz}
(the most general tractable one) is the Gaussian form
\begin{gather} \label{shoans2}
\rho[A,A^{'}] = \exp{-\frac{1}{2} [A G^{-1} A + A^{'} G^{-1} A^{'} - 2 A H A^{'}]},
\end{gather}
where we employ a matrix notation for integrals
\begin{gather} \label{matnot}
A G H A = \int dxdydz \; A(x) G (x-y) H (y-z) A(z) 
\end{gather}
\emph{etc.\ }and the functions $G$ and $H$ are arbitrary kernels with respect to which the free energy will be minimised.

The Helmholtz free energy is $F = U - TS$ where $U$ is the energy and
$S$ is the entropy. Remembering that $\rho$ is no longer normalised, one has
\begin{gather} \label{shoened}
U = \frac{\mathrm{tr} \rho \mathcal{H}}{\mathrm{tr} \rho}
\end{gather}
and
\begin{gather} \label{shoentd}
S = -\frac{\mathrm{tr} [ \rho \log(\rho /\mathrm{tr} \rho) ]}{\mathrm{tr} \rho} 
= - \frac{\mathrm{tr} \rho \log{\rho}}{\mathrm{tr} \rho} + \log \mathrm{tr} \rho.
\end{gather}
Here, $\mathcal{H}$ is the Hamiltonian
\begin{gather}
\mathcal{H} [A,A^{'}]= \frac{1}{2} \delta(A - A^{'}) \left[ -\frac{\delta^2}{\delta A^2} + A \omega^2 A \right] .
\end{gather}
The ensemble average of the energy (\ref{shoened}) with respect to (\ref{shoans}) is a simple Gaussian integral. One obtains
\begin{align}
U &= \frac{1}{2\mathrm{tr} \rho} \int DADA^{'} \; 
\delta(A - A^{'}) \left[ -\frac{\delta^2}{\delta A^2} + A \omega^2 A \right] \rho[A,A'] \nonumber \\
 &=  \frac{1}{2\mathrm{tr} \rho} \int DA \; 
 [ \mathrm{tr} G^{-1} - A(G^{-1} - H)^2A + A \omega^2 A] \rho[A,A] \nonumber \\
 &= \frac{1}{4} \mathrm{tr} (G^{-1} + H + \omega^2 (G^{-1} - H)^{-1} ).
\end{align}
In momentum space
\begin{gather} \label{shoenec}
U = \frac{V}{4} \int \frac{dp}{2\pi} \left[ G^{-1} + H + \omega^2 (G^{-1} - H)^{-1} \right],
\end{gather}
where V is the spatial volume.
How to evaluate the entropy (\ref{shoentd}) in general is less clear. Indeed it is not apparent how to \emph{define} the 
logarithm of $\rho$. However,
for a Gaussian density matrix we can use the harmonic oscillator correspondence proven in the previous section. 
In terms of $(\omega,T)$, the entropy is given by
\begin{gather}
S = \frac{\partial}{\partial T} T\log{Z} = - \log (1-e^{-\omega /T}) +\frac{\omega e^{-\omega /T}}{T(1-e^{-\omega /T})},
\end{gather}
which can be re-written in terms of $G$ and $H$ (in fact in terms of just $h = GH/2$) via (\ref{shokers}) as
\begin{gather} \label{shoentk}
S =  + \log \left[ \frac{2h}{ (1-4h^2)^{1/2} - (1-2h)}\right] 
- \log \left[ \frac{1- (1-4h^2)^{1/2}}{2h}\right] 
\left( \frac{1- (1-4h^2)^{1/2}}{\left( (1-4h^2)^{1/2} - (1-2h)\right)} \right).
\end{gather}
 The free energy in terms of $G$ and $h$ is, from (\ref{shoenec}) and (\ref{shoentk}),
\begin{multline}
F = \frac{1}{4}\left[ G^{-1}(1 + 2h) + \omega^2 G(1 - 2h)^{-1}\right] \\
- T \left(  \log \left[ \frac{2h}{  (1-4h^2)^{1/2} - (1-2h)}\right] 
- \log \left[ \frac{1- (1-4h^2)^{1/2}}{2h}\right] \left( \frac{1- (1-4h^2)^{1/2}}{(1-4h^2)^{1/2}-(1-2h)} \right) \right).
\end{multline}
Minimising with respect to $G$, one obtains
\begin{gather}
G^{-1} = \omega (1- 4h^2)^{-1/2}.
\end{gather}
Minimising with respect to $h$, one obtains
\begin{gather}
0 = G^{-1} + \omega^2 G (1-2h)^{-2} + \frac{2T}{(1+ 2h)^{1/2} (1-2h)^{3/2}} \log \left[ \frac{1- (1-4h^2)^{1/2}}{2h} \right].
\end{gather}
Substituting the former expression for $G$ into the latter expression 
yields the solutions
\begin{align} \label{shosoln}
G^{-1} &= \omega \frac{1+e^{-2\omega / T}}{1-e^{-2\omega / T}} \nonumber \\
H &= 2 \omega \frac{e^{-\omega / T}}{1-e^{-2\omega / T}},
\end{align}
which are the kernels derived in section \ref{sec:sho} (\ref{shokers}).
\section{Compact U(1) gauge theory} \label{sec:cQED}
Gauge theories present an added complication: it must be guaranteed that gauge invariance is respected throughout.
So physical observables must be gauge invariant and furthermore we must only consider states that belong
to the gauge-invariant sector of the Hilbert space. 
For the variational method at $T=0$ this is achieved by integrating over the gauge group in the \emph{ansatz} for
the wavefunction to project out gauge-invariant states \cite{Kogan:1995wf}.
Correspondingly, in (\ref{shodens}) we should restrict the sum to
gauge-invariant states, so we project the \emph{ansatz} onto the gauge-invariant sector. 
Because the density matrix is bi-linear in the states, this requires \emph{two} gauge projections.

Consider a theory of a gauge field $A_{\mu}$ with gauge symmetry under a gauge transformation $\varphi$ such that 
$A_{\mu} \rightarrow A_{\mu}^{\varphi}$. Working in the Hamiltonian formalism of the field theory with $A_0=0$, the required gauge invariance
can thus be achieved by modifying the density matrix \emph{ansatz} (\ref{shoans2}) to
\begin{gather} \label{nqedans1}
\rho^{'} [A,A^{'}] = \int D\varphi D\varphi' \; 
\exp{-\frac{1}{2} 
\left[ A_{i}^{\varphi} G^{-1} A_{i}^{\varphi} + A_{i}^{'\varphi'} G^{-1} A_{i}^{'\varphi'} - 2 A_{i}^{\varphi} H A_{i}^{'\varphi'} \right]}.
\end{gather}
Here $i$ labels the spatial components of the gauge field.
But if operator $O[A,A']$ is gauge invariant, one of the gauge group integrals in $\mathrm{tr} \rho O$ is redundant. It suffices to take
\begin{gather} \label{nqedans}
\rho [A,A^{'}] = \int D\varphi \; 
\exp{-\frac{1}{2} \left[ A_{i} G^{-1} A_{i} + A_{i}^{'\varphi} G^{-1} A_{i}^{'\varphi} - 2 A_{i} H A_{i}^{'\varphi} \right] }.
\end{gather}

Consider now applying the variational method to \emph{non-compact} $U(1)$ in $2+1$ dimensions. 
The relevant $U(1)$ gauge transformation is $A_{i} \rightarrow A_{i}^{\varphi} = A_{i} - \partial_i \varphi / g$,
where $g$ is the coupling and $\varphi$ is a \emph{regular} function.
We could now follow the variational method recipe given above verbatim: 
determine the energy with respect to the density matrix \emph{ansatz} (\ref{nqedans}) and form the free energy $F$,
including the entropy  of a single massless degree of freedom.\footnote{Entropically speaking, free photons in $2+1$ dimensions
with a single transverse degree of freedom are equivalent to a single massless scalar degree of freedom.} 
Then minimise $F$ with respect to the variational kernels $G$ and $h$. 
We choose not to do so here, because the result can be recovered from the \emph{compact} theory discussed below by letting
the coupling go to zero (in which case the magnetic monopoles decouple from the photons).
One finds that the kernels are the same as for the single free massless scalar field, as one expects.

The foregoing discussion has given us a detailed understanding of the density matrix and finite temperature variational method
for what are essentially trivial free theories. It is now time to consider a less trivial theory: the compact $U(1)$ gauge theory.
In the Hamiltonian formalism, 
the compact theory differs from the non-compact theory in that it admits singular gauge transformations $\varphi (x)$ 
which create vortex-like gauge field configurations
which act in the partition function as magnetic monopoles.
At zero temperature these monopoles form a dilute plasma, 
causing Debye screening of the $U(1)$ field. Thus the photon has a mass gap, 
the Wilson loop exhibits an area law and the theory is confining.

An important point to note is that the compact $U(1)$ theory is in a sense very similar at weak coupling 
to the non-compact $U(1)$ theory: the monopole plasma is extremely dilute, and the photon mass is non-perturbatively small.
Thus one expects that the Gaussian variational \emph{ansatz}
(\ref{nqedans}) will be a suitable one. 

The variational method at finite temperature followed below very closely parallels the analysis at zero temperature given in 
\cite{Kogan:1995vb}, to which the reader should refer for details. The principle is essentially the same, 
but the presence of the additional kernel $H$ and entropic effects complicate the algebra.

We start with a short discussion of thermal ensemble averages. Consider first the simplest of all \emph{viz.\ }$\mathrm{tr} \rho$.
This is of course just the normalization. Now 
\begin{align} \label{action}
\mathrm{tr} \rho &= \int D\varphi DA_i \; 
\exp{ -\frac{1}{2} [A_i G^{-1} A_i + A_{i}^{\varphi} G^{-1} A_{i}^{\varphi} - 2A_i H A_{i}^{\varphi}]} \nonumber \\
&= \int D\varphi DA_i \;
\exp{ - \left[ A_{i}^{\varphi /2} (G^{-1} - H) A_{i}^{\varphi /2} + \frac{1}{4g^2} \partial_i \varphi (G^{-1} + H) \partial_i \varphi \right]} \nonumber \\
&= \int D\varphi DA_i \; \exp{ - W }.
\end{align}
As we shall see below, ensemble averages translate into the language of field theory as correlation functions
evaluated with respect to Euclidean action $W$.

To continue the evaluation of (\ref{action}), split the gauge transformation $\varphi$ into parts $\tilde{\varphi}$ and $\varphi_v$,
corresponding respectively to regular (non-compact) gauge transformations and to a distribution of $n_+$ vortices at points $x_{\alpha}$
and $n_-$ anti-vortices at points $x_{\beta}$. Explicitly,
\begin{gather}
\varphi_v = \Sigma_{\alpha = 1}^{n_{+}} n \theta (x - x_{\alpha}) - \Sigma_{\beta = 1}^{n_{-}} n \theta (x - x_{\beta}),
\end{gather}
where $\theta$ is the plane polar angle and $n$ is the charge (in units of $2\pi/g$) on a fundamental monopole. 
Then the normalisation integral (\ref{action}) can be evaluated as follows.
The integrals over $A_i$ and $\tilde{\varphi}$ are Gaussian. Further, substituting the general vortex distribution for $\varphi_v$
we can write the $\varphi_v$ integral in standard fashion as the partition function $Z_v$ for a Coulomb gas. In all,
\begin{gather}
\mathrm{tr} \rho = \det \left[ \pi (G^{-1} - H)^{-1} \right] \; \det \left[ \frac{4 \pi g^2}{\partial^2} (G^{-1} + H)^{-1} \right]^{1/2}\; Z_v,
\end{gather}
where
\begin{multline}
Z_v = \Sigma_{n_{+} , n_{-} =0}^{\infty} \frac{1}{n_{+} ! n_{-} !} 
\Pi_{\alpha = 1}^{n_{+}} \Pi_{\beta = 1}^{n_{-}}
\int d^2 x_{\alpha} d^2 x_{\beta} \; z^{n_{+} + n_{-}}  \\
 \exp {-\frac{1}{4g^2} \left[ \Sigma_{\alpha, \alpha^{'}}  D(x_{\alpha} - x_{\alpha^{'}}, G^{-1} + H ]  
+ \Sigma_{\beta, \beta^{'}}  D(x_{\beta} - x_{\beta^{'}}, G^{-1} + H ]  
- \Sigma_{\alpha, \beta}   D(x_{\alpha} - x_{\beta}, G^{-1} + H ] \right]}.
\end{multline}
In this last expression for the vortex partition function, the vortex-vortex interaction is\footnote{The notation
$D(x, G^{-1} + H ]$ indicates that $D$ is a function of the separation $x$ and a functional of the kernel $G^{-1} + H$.}
\begin{gather} \label{vorint}
D(x, G^{-1} + H ] = 8 \pi^2 n^2 \int_{0}^{\Lambda} \frac{d^2 p}{(2 \pi)^2} \; \frac{G^{-1} + H}{p^2} e^{ipx},
\end{gather}
the vortex fugacity is
\begin{gather} \label{fug}
z = \Lambda^2 \exp{\left[ - \frac{1}{8 g^2} D (0 , G^{-1} + H] \right] },
\end{gather}
and $\Lambda$ is the UV cutoff in the theory.

Note that, since the minimal kernels $G$ and $H$ will in general have temperature dependence, the monopole interaction $D$
and fugacity $z$ will also acquire $T$ dependence.
This is in accord with the dimensional reduction arguments of \cite{Agasian:1998wv}.

After this brief diversion, let us get back to the discussion of thermal ensemble averages.
Ensemble averages translate into field-theoretic language as correlation functions with respect 
to the `action' $W$ in (\ref{action}). This action has three fields: $A_i$, $\tilde{\varphi}$ and $\varphi_v$. The former two are free
fields. Using the matrix notation of (\ref{matnot}) and ignoring normalization factors (these cancel in practice), one can
immediately write down the following correlators:
\begin{align}
\langle A_i \rangle_A &= \frac{1}{2g} \partial_i \varphi,\\
\langle A_i A_i \rangle_A &= (G^{-1} - H)^{-1} + \frac{1 }{4g^2} \partial_i \varphi \partial_i \varphi,\\
\langle \partial_i \tilde{\varphi} \rangle_{\tilde{\varphi}} &= 0,\\
\langle \partial_i \tilde{\varphi} \partial_i \tilde{\varphi} \rangle_{\tilde{\varphi}} &= 2  g^2 (G^{-1} + H)^{-1}.
\end{align}
The vortex field correlators are evaluated as follows. We write them as correlation functions of the vortex density
\begin{gather}
\rho (x) = \Sigma_{\alpha \beta} n [\delta (x - x_{\alpha }) - \delta (x - x_{\beta})].
\end{gather}
The two-point vortex density correlator is identical to that in \cite{Kogan:1995vb}, 
if one takes account of the modified form of $z$ in (\ref{fug}) . The result is
\begin{gather}
\langle \rho \rho \rangle_{\rho} = 2z +O(z^2).
\end{gather}

These results are all that is needed to determine the energy with respect to the \emph{ansatz} (\ref{nqedans}). 
The Hamiltonian for compact $U(1)$  is
\begin{gather}
\mathcal{H}  = \frac{1}{2} [E_{i}^{2} + b^2].
\end{gather}
As \emph{per} the Hamiltonian for non-compact $U(1)$, the electric contribution to the energy is 
\begin{gather}
\frac{1}{2} E_{i}^{2} = - \frac{1}{2} \delta (A - A') \frac{\delta^2}{\delta A_{i}^{2}}.
\end{gather}
However, the magnetic part is \emph{not} the usual
\begin{gather}
\frac{1}{2} B^{2} = \frac{1}{2} \delta (A - A') (\epsilon_{ij} \partial_i A_j )^2.
\end{gather}
This is because $B$ is \emph{not} invariant under compact gauge transformations, 
which in general create sources of magnetic flux, \emph{i.e.\ }the monopoles. We require a singlet part of $B^2$ which is invariant
under compact gauge transformations and which we denote by $b^2$.

The normalised electric contribution to the energy in the thermal ensemble is thus
\begin{gather}
\frac{\mathrm{tr} \rho \frac{1}{2} E_{i}^{2}}{\mathrm{tr} \rho} =  - \frac{1}{2 \mathrm{tr} \rho} \int DA DA' \;
\delta (A - A') \frac{\delta^2}{\delta A_{i}^{2}} \rho [A,A'].
\end{gather}
Doing the functional differentiation and performing the $A_{i}$, $\tilde{\varphi}$ and $\varphi_v$ integrations in turn, this is
\begin{align}
& \frac{1}{2 \mathrm{tr} \rho} \int DA D\varphi \;
\delta (A - A') \left[\delta_{ii} \mathrm{tr} G^{-1} - (A_i G^{-1}- A_{i}^{' \varphi} H )(G^{-1} A_i - H A_{i}^{' \varphi}) \right] 
e^{-W} \nonumber \\
= &\frac{1}{2 \mathrm{tr} \rho} \int DA D\varphi \;
\left[ \mathrm{tr} (G^{-1} + H) - \frac{1}{4g^2} \partial_i \varphi (G^{-1} + H )^2 \partial_i \varphi \right] 
e^{-W} \nonumber \\
= &\frac{1}{2 \mathrm{tr} \rho}  \int DA D\tilde{\varphi} D \tilde{\varphi} \; 
\left[ \frac{1}{2} \mathrm{tr}(G^{-1} + H) - \frac{1}{4g^2} \partial_i \varphi_v (G^{-1} + H )^2 \partial_i \varphi_v \right] 
e^{-W} \nonumber \\
= &\frac{1}{4}
\left[\mathrm{tr} (G^{-1} + H) - \frac{1}{4g^2} \langle \rho D[ (G^{-1} + H )^2 ]\rho \rangle_{\rho} \right],
\end{align}
where $D[ (G^{-1} + H )^2 ]$ is the vortex interaction for kernel $(G^{-1} + H )^2$, \emph{cf.\ }(\ref{vorint}).
In momentum space, using $\langle \rho \rho \rangle_{\rho} = 2z$, one gets
\begin{gather} \label{elec}
\frac{\mathrm{tr} \rho \frac{1}{2} E_{i}^{2}}{\mathrm{tr} \rho} =
V \int \frac{d^2 p}{(2 \pi)^2} \left[ \frac{1}{4}(G^{-1} + H) - \frac{ n^2 \pi^2 z}{g^2 p^2} (G^{-1} + H )^2 \right].
\end{gather}
This has the same form as in the $T=0$ calculation with the replacement $G^{-1} \rightarrow G^{-1}+H$; the difference
is that $G,H$ and $z$ will acquire $T$ dependence upon minimisation at non-zero temperature. 

Next consider the magnetic contribution to the energy. The compact $U(1)$ Hamiltonian contains $b^2$, a part of $B^2$ which is
a singlet under compact gauge transformations.
The existing literature on the Hamiltonian treatment of compact $U(1)$ at $T=0$ suggests two ways in which to
proceed here. The first \cite{Kogan:1995vb} is to note that
the ensemble averages of $\mathrm{tr} \rho b^2 = \mathrm{tr} \rho B^2$ should coincide
when taken with respect to the properly compact gauge-invariant density matrix, so it suffices to calculate the latter. 
There is a \emph{caveat} however: 
because $B^2$ is not compact gauge invariant, we must use (\ref{nqedans1}) rather than (\ref{nqedans}) as the \emph{ansatz}
for the density matrix, since the latter form is only valid for operators which are fully gauge-invariant.

The second method is to explicitly choose a singlet form for $b^2$. In \cite{Kovner:1998eg}, the authors chose
the `mixed action'
\begin{gather}
b^2 = -\frac{1}{g^2} \int \left[ \alpha \cos gB + \frac{(1-\alpha)}{4} \cos 2gB \right],
\end{gather}
which reduces at weak coupling to the standard non-compact magnetic term. At zero temperature, this theory contains a surprise:
for $\alpha > \pi^2/4$ there is no mass gap in the best variational state. 
So the different methods give different results at zero-temperature, presumably because the singlet parts are different.
However, we note (but do not show here)
that at finite temperature, the analysis of the theory with mixed action parallels the first method above: if there is a mass
gap at $T=0$ ($\alpha < \pi^2/4$), it will disappear at $T_c$; if there is no mass gap at $T=0$, there is no mass gap anywhere.

Following the dogma of the first method, one has that
\begin{align}
\frac {\mathrm{tr} \rho \frac{b^2}{2}}{\mathrm{tr} \rho} &= 
\frac {\mathrm{tr} \rho^{'} \frac{B^2}{2}}{\mathrm{tr} \rho^{'}} \nonumber \\
&= \frac {1}{2\mathrm{tr} \rho^{'}} \int DA DA^{'} D \varphi' D \varphi''\; B^2
\exp{-\frac{1}{2} \left[ A_{i}^{\varphi^{'}} G^{-1} A_{i}^{\varphi^{'}} + A_{i}^{'\varphi^{''}} G^{-1} A_{i}^{'\varphi^{''}} 
- 2 A_{i}^{\varphi^{'}} H A_{i}^{'\varphi^{''}} \right] }. 
\end{align}
Now make the change of variables $A,A^{'} \rightarrow A^{-\varphi'} , A^{'-\varphi'}$ and let 
$\varphi = \varphi^{''} - \varphi^{'}$, $\eta = \varphi^{''} + \varphi^{'}$ such that
\begin{gather}
\frac {\mathrm{tr} \rho \frac{b^2}{2}}{\mathrm{tr} \rho} =
\frac {1}{2\mathrm{tr} \rho^{'}} \int DA D \varphi D \eta \;
\left[ \epsilon_{ij} \partial_i \left(A_j - \frac{1}{2g} \partial_j (\varphi - \eta)\right) \right]^2 e^{-W}.
\end{gather}
The term linear in $\eta$ vanishes, and the quadratic term is independent of the variational kernels; we omit it. Then
\begin{gather}
\frac {\mathrm{tr} \rho \frac{b^2}{2}}{\mathrm{tr} \rho} =
\frac {1}{2\mathrm{tr} \rho^{'}} \int DA D \varphi D \eta \;
\left[ \epsilon_{ij} \partial_i \left( A_j - \frac{1}{2g} \partial_j \varphi \right) \right]^2 e^{-W}.
\end{gather}
Making the change of variables $A \rightarrow A^{-\varphi /2 }$ yields
\begin{gather} \label{mag}
\frac {\mathrm{tr} \rho \frac{b^2}{2}}{\mathrm{tr} \rho} = \frac{1}{4} \mathrm{tr} \; \partial_{i}^{2} (G^{-1} - H)^{-1}.
\end{gather}
Altogether, the ensemble average of the internal energy, the sum of electric (\ref{elec}) and magnetic (\ref{mag}) parts, is
\begin{gather}
U = V \int \frac{d^2 p }{4 (2 \pi)^2} \left[ G^{-1} + H - \frac{4 n^2 \pi^2 z}{ g^2 p^2} (G^{-1} + H )^2 + p^2 (G^{-1} - H)^{-1} \right].
\end{gather}

Again it is unclear how to determine the entropy from first principles.\footnote{Because of the non-trivial gauge projection,
the density matrix is no longer Gaussian.} 
We conjecture that the entropy \emph{written in terms of the kernels}
is the same as for a free  theory, as in (\ref{shoentk}). Why should this be the case? We expect to find a solution
in which the photon is still approximately free, but has a mass gap. If this is the case,
then entropically speaking the compact photon corresponds to a single free \emph{massive} degree of freedom.
But this possibility is contained in the expression for the entropy written in terms of the arbitrary kernels. 
Turning this argument upon its head, we can justify the conjecture \emph{a posteriori} in the following fashion.
Let us make the conjecture, carry out the minimisation and ask the question:
Is the entropy thus obtained that of a single massive degree of freedom? As we shall see, it is.

Putting everything together, the free energy in terms of $G$ and $h$ to order $z$ is
\begin{multline}
F = \frac{V}{4} \int \frac{d^2p}{(2\pi)^2} 
\left[  G^{-1}(1 +2h) + p^2 G (1 -2h)^{-1} 
- \frac{4 n^2 \pi^2 z (1+2h)^2}{g^2 p^2 G^2} \right. \\
\left. - T \left(  \log \left[ \frac{2h}{  (1-4h^2)^{1/2} - (1-2h)}\right] 
- \log \left[ \frac{1- (1-4h^2)^{1/2}}{2h}\right] \left( \frac{1- (1-4h^2)^{1/2}}{(1-4h^2)^{1/2}-(1-2h)} \right) \right) \right].
\end{multline}

Minimising with respect to $G$, one obtains
\begin{gather} \label{gmin}
0 = - G^{-2}(1 +2h) + p^2 (1 -2h)^{-1} 
- \frac{4 n^2 \pi^2 }{g^2}\frac{\delta z}{\delta G(p)} \int d^2k \frac{(1+2h)^2}{k^2 G^2} 
+ \frac{8 n^2 \pi^2 z (1+2h)^2}{g^2 p^2 G^3}. 
\end{gather}
Minimising with respect to $h$, one obtains
\begin{multline} \label{hmin}
0 = G^{-1} + p^2 G (1-2h)^{-2}
- \frac{2 n^2 \pi^2 }{g^2}\frac{\delta z}{\delta h(p)} \int d^2k \frac{(1+2h)^2}{k^2 G^2}
+ \frac{8 n^2 \pi^2 z (1+2h)}{g^2 p^2 G^2} \\
+ \frac{2T}{(1+2h)^{1/2} (1-2h)^{3/2}} \log \left[ \frac{1- (1-4h^2)^{1/2}}{2h} \right].
\end{multline}
From (\ref{fug}) one has the following relations
\begin{gather}
\frac{\delta z}{\delta h } = - \frac{n^2 z}{2 g^2 p^2 G },\\ \frac{\delta z}{ \delta G } = \frac{n^2 z (1+2h) }{4 g^2 p^2 G^2}.
\end{gather}
We are interested in the weak coupling regime, where $\Lambda /g^2 \gg 1$, and the low temperature regime, 
where $0 \leq T \lesssim g^2$. If one evaluates the $z$ terms in the minimisation equations (\ref{gmin},\ref{hmin}) in this regime,
one finds that terms with $\frac{\delta z}{\delta G,h }$ dominate over the others, which we discard.
Thus, substituting for $G$ from (\ref{gmin}) into (\ref{hmin}), one has
\begin{gather}
\frac{1 - (1-4h^2)^{1/2}}{2h} = e^{-\frac{(p^2 + m^2)^{1/2}}{T}},
\end{gather}
which is the Boltzmann factor for a massive photon, where the mass is given by
\begin{gather} \label{mass}
m^2 =   \frac{n^4 \pi^2  z}{g^4} \int d^2k \; \frac{(1+2h)^2}{k^2 G^2} \simeq \frac{n^4 \pi^3 \Lambda^2 z}{g^4}.
\end{gather}
Here we we have used the approximations $h \simeq 0$ and $G^{-1} \simeq k$ for large $k$.
The form of the Boltzmann factor ensures that the entropy (which depends on $h$ but not $G$) corresponds to the entropy
of a single relativistic massive excitation of mass $m$. Thus our previous conjecture that compact $U(1)$ is (entropically speaking)
a free theory is verified.

It is now straightforward to solve for the kernels. One obtains
\begin{align} \label{kers}
G^{-1} &= \frac{p^2}{(p^2 +m^2)^{1/2}} 
\left( \frac{1+ e^{-\frac{2(p^2 +m^2)^{1/2}}{T}}}{1 - e^{-\frac{2(p^2 +m^2)^{1/2}}{T}}}\right), \nonumber \\
H &= 2\frac{p^2}{(p^2 +m^2)^{1/2}}
 \left( \frac{e^{-\frac{(p^2 +m^2)^{1/2}}{T}}}{1 - e^{-\frac{2(p^2 +m^2)^{1/2}}{T}}}\right).
\end{align}
In the limit as $g^2/\Lambda \rightarrow 0$, $ m \rightarrow 0$ and one obtains the expected
kernels for the non-compact theory, \emph{cf.\ }(\ref{shokers}) with $\omega=p$.
\section{Phase Structure of Compact $U(1)$} \label{sec:phase}
It is now straightforward to analyse the phase structure of the theory. 
From equations (\ref{vorint}) and (\ref{fug}) the fugacity is given by
\begin{gather}
z = \Lambda^2 \exp{- \frac{n^2 \pi}{2 g^2} \int_{0}^{\Lambda} pdp \; \frac{G^{-1} + H}{p^2} }.
\end{gather}
Now substitute for the kernels $G$ and $H$ from (\ref{kers}) to obtain
\begin{gather}
z = \Lambda^2 \exp{- \frac{n^2 \pi}{2 g^2} \int_{0}^{\Lambda} pdp \; \frac{1}{(p^2+m^2)^{1/2}}
\coth \frac{(p^2+m^2)^{1/2}}{2T}}.
\end{gather}
The integral is straightforward. One obtains
\begin{align}
z &= \Lambda^2 \exp{- \frac{n^2 \pi}{2 g^2} \left[ 2T \log \sinh \frac{(p^2 + m^2)^{1/2}}{2T}\right]_{0}^{\Lambda}},
\nonumber \\
&\simeq \Lambda^2 \exp{- \frac{n^2 \pi}{2 g^2}\left[  \Lambda - 2T \log m/T \right]}, \nonumber \\
&= \Lambda^2 \exp{- \frac{n^2 \pi \Lambda}{2 g^2}} \left( \frac{m^2}{T^2} \right)^{\frac{n^2 \pi T}{2 g^2}}.
\end{align}Using the expression (\ref{mass}) for $m$ yields a self-consistent equation for the dimensionless
fugacity $\tilde{z} = z/\Lambda^2$:
\begin{gather} \label{fugfug}
\tilde{z} = e^{- \frac{n^2 \pi \Lambda}{2 g^2}} 
\left( \frac{n^4 \pi^3 \Lambda^4 \tilde{z}}{g^4 T^2} \right)^{\frac{n^2 \pi T}{2 g^2}},
\end{gather}
with the two solutions
\begin{align} \label{zlh}
\tilde{z} &= 0, \nonumber \\
\tilde{z}^{(1 - \frac{n^2 \pi T}{2 g^2})} &= e^{- \frac{n^2 \pi \Lambda}{2 g^2}} 
\left( \frac{n^4 \pi^3 \Lambda^4}{g^4 T^2} \right)^{\frac{n^2 \pi T}{2 g^2}}.
\end{align}
The first solution is simple to comprehend. What about the second? The right hand side
is, for $0 < T \lesssim g^2$, a non-perturbatively small dimensionless quantity. At $T=0$,
one recovers the Kogan and Kovner result for the fugacity \cite{Kogan:1995vb}. As $T$ increases from zero,
the left hand side is $\tilde{z}$ raised to a fractional power. This implies that the value of $\tilde{z}$
which satisifes the equation is smaller and indeed goes continuously to zero as $T$ goes to $2 g^2/n^2\pi$.
More explicitly,
writing $t = n^2 \pi T/2 g^2$ and taking the logarithm, 
one has that
\begin{gather}
\log \tilde{z} = - \frac{f(t)}{1-t},
\end{gather}
where $f(t)$ is a function of $O(\Lambda/g^2)$ which is both positive and smooth ($C^{\infty}$) at $t=1$. Thus,
\begin{gather} \label{tilz}
\tilde{z} = \exp{- \frac{f(t)}{1-t}}.
\end{gather}
As $t$ increases towards unity, $\tilde{z}$ goes continuously to zero; above $t=1$, $\tilde{z}$ becomes large.

What are we to make of all this? We have obtained two solutions which \emph{extremise} the free energy. 
It is clear that the second solution \emph{minimises} the free energy for temperatures below $2 g^2/n^2\pi$
(otherwise there would be no mass gap at zero temperature), but is unphysical (we only work to $O(z)$ remember)
for temperatures above $2 g^2/n^2\pi$ and should be discarded. So at high temperatures we are left only with the solution $z=0$.

Hence there is a critical phase transition at 
\begin{gather} \label{T_c}
T_c = 2 g^2/n^2\pi, 
\end{gather}
from a phase with non-zero fugacity and mass gap
(corresponding to an unbound plasma of monopoles and confinement of electric charges)
 to a phase with zero fugacity and mass gap
(corresponding to bound monopole-antimonopole pairs and deconfinement). 

Can we expand on the nature of this phase transition? Indeed we can. Consider again the low temperature solution (\ref{tilz})
for $\tilde{z}$. Not only does $\tilde{z}$ go to zero as $t$ tends to unity from below, but so also do all derivatives
of $\tilde{z}$ with respect to $t$, because $f$ is smooth and the exponential falls off faster than any power of $(1-t)$. For example,
\begin{gather}
\lim_{t \rightarrow 1} \frac{dz}{dt} = \lim_{t \rightarrow 1} e^{- \frac{f(t)}{1-t}} \left[ \frac{f'}{1-t} - \frac{f}{(1-t)^2} \right] = 0,
\end{gather}
and so on.
Thus we have that $\tilde{z}$ and all its derivatives vanish on the low temperature side of the transition,
as indeed they do on the high temperature side of transition (where $\tilde{z}=0$).
Thus $\tilde{z}$ and all its derivatives are \emph{continuous} at the transition.
Moreover,
this smoothness of $\tilde{z}$ extends trivially to the free energy $F$, since any pathologies
of $F$ at $t=1$ can only come from $\tilde{z}$. Thus, finally, we see that the free energy and all its derivatives
are continuous at the transition:
\begin{gather} \label{cont}
\frac{d^m F}{d t^m} = 0.
\end{gather}
Thus, the transition is of infinite order according
to the Ehrenfest classification. Since we do not know of another univerality class which has this property, we
suggest that the phase transition does indeed belong to the BKT universality class, as expected.
\section{Correlation Functions and the Wilson Loop} \label{sec:corrs}
Having obtained the minimal variational kernels (\ref{kers}), it is now possible to calculate correlation functions in
the compact $U(1)$ theory. For example, the magnetic field propagator is given by (following the arguments preceeding
eq'n \ref{mag})
\begin{gather}
\langle b(x) b(y) \rangle =
\frac {1}{2\mathrm{tr} \rho^{'}} \int DA D \varphi D \eta \;
\left[ \epsilon_{ij} \partial_i \left(A_j - \frac{1}{2g} \partial_j (\varphi - \eta)\right) \right]_x
\left[ \epsilon_{kl} \partial_k \left(A_l - \frac{1}{2g} \partial_l (\varphi - \eta)\right) \right]_y
e^{-W},
\end{gather}
where all quantities in the first set of parentheses take the argument $x$ and so on. Hence,
\begin{gather}
\langle b(x) b(y) \rangle = \frac{1}{2} \partial_{i}^{x} \partial_{i}^{y} (G^{-1} - H )^{-1} (x-y),
\end{gather}
or in momentum space
\begin{gather}
\langle b(p) b(-p) \rangle = \frac{ (p^2 + m^2)^{1/2}}{2} \coth \frac{(p^2 + m^2)^{1/2}}{2T}.
\end{gather}
This is to be compared with the magnetic propagator for the non-compact theory 
at finite temperature calculated in the standard Lagrangian formalism.
There (in Feynman gauge and neglecting decoupled ghosts) the Euclidean effective action is
\cite{Kapusta:1989tk}
\begin{gather}
\int_{0}^{\beta} d \tau \int d^2x \; \frac{1}{2} A_{\mu} \left( \partial^{2}_{\tau} + \partial^{2}_{x}\right) A_{\mu},
\end{gather}
obtained from the Minkowski action via the replacements $t \rightarrow -i\tau$ and $A_0 \rightarrow i A_0$.
The magnetic propagator in terms of the Matsubara frequencies $\omega_n = 2 \pi n T$ is then
\begin{gather} \label{lag}
\langle B(x,\tau) B(y, \tau) \rangle = 
\Sigma_{n=-\infty}^{\infty} \int \frac{d^2 p}{(2 \pi)^2} \; \frac{p^2}{\beta (p^2 + \omega_{n}^{2})} e^{ip(x-y)}
= \int \frac{d^2 p}{(2 \pi)^2} \; \frac{ p}{2} \coth \frac{ p}{2T}.
\end{gather}

In the low temperature limit, the compact magnetic field propagator is simply $(p^2 + m^2)^{1/2}/2$ and taking the Fourier transform
(see \emph{e.g.\ }\cite{Gripaios:2002bu}) one sees that the correlation functions decay as $e^{-m|x-y|}$ at large separations.
In the high temperature limit, one has
\begin{gather}
\langle b(p) b(-p) \rangle \rightarrow T,
\end{gather}
or in co-ordinate space
\begin{gather}
\langle b(x) b(y) \rangle \rightarrow T \delta (x-y).
\end{gather}
This too is to be expected: at high temperatures, 
fields undergo large thermal fluctuations but are uncorrelated at distinct space points.

Strictly speaking, the correct propagator to consider is
\begin{multline}
\langle b^2(x) b^2(y) \rangle = \\
\frac {1}{2\mathrm{tr} \rho^{'}} \int DA D \varphi D \eta \;
\left[ \epsilon_{ij} \partial_i \left(A_j - \frac{1}{2g} \partial_j (\varphi - \eta)\right) \right]^{2}_{x}
\left[ \epsilon_{kl} \partial_k \left(A_l - \frac{1}{2g} \partial_l (\varphi - \eta)\right) \right]^{2}_{y}
e^{-W},
\end{multline}
since it is $b^2$ that is invariant under the full set of compact $U(1)$ gauge transformations.
There is a complication here in that there are terms containing $\eta$ which appear to contribute to the propagator
yet belong to an ill-defined path integral. The only $\eta$ configurations which contribute to the path integral
are those which are singular, since $\epsilon_{ij} \partial_i \partial_j \eta = 0$ when $\eta$ is regular.
But $\eta$ only features in the path integral in the combination $\varphi - \eta$; if we allow $\varphi$
to be singular but insist that $\eta$ be regular, we generate all possible configurations
$\varphi - \eta$ (singular or otherwise)
and at the same time ensure that the path integral is well defined. Via the Wick expansion, one obtains
\begin{gather}
\langle b^2(x) b^2(y) \rangle = 2 \langle b(x) b(y) \rangle^2,
\end{gather}
which for small $T$ falls off as $e^{-2m|x-y| }$ at large separations.

Lastly, let us consider the spatial Wilson loop written in the form
\begin{gather}
W_{\partial S} = \langle e^{ilg \oint_{\partial S} A_i d x_i } \rangle = 
\langle  e^{ilg \int_{S} B d S } \rangle,
\end{gather}
where $l$ takes integral values, ensuring that the Wilson loop is gauge invariant under compact transformations of the
form $B \rightarrow B + 2\pi /g$. Then
\begin{multline} 
W_{\partial S} =  
\int DA_i \; \exp{ \left[ -A (G^{-1} - H) A + ilg \int_{S} \epsilon_{ij} \partial_i A_j d^2 x \right]} \\
\times \int D \varphi \; 
\exp{ \left[ -\frac{1}{4g^2} \partial_i \varphi (G^{-1} + H) \partial_i \varphi 
+ \frac{il}{2} \oint_{\partial S} \partial_i \varphi d x_i \right]}.
\end{multline} 
For a loop of large area, the first term is
\begin{align} 
  &\exp{-\frac{l^2 g^2}{2} \int_S \langle B(x) B(y) \rangle d^2x d^2y} \nonumber \\
= &\exp{-\frac{l^2 g^2}{2} S \lim_{p \rightarrow 0} \frac{p^2}{2} (G^{-1} - H )^{-1}} \nonumber \\
= &\exp{-\frac{l^2 g^2}{4} S m \coth \frac{m}{2T}}. 
\end{align}
The second term gives a vortex contribution of $O(z) \simeq O(m^2)$ which is sub-dominant. So we have the apparently curious
result that the area law behaviour persists for all $T$. Indeed,
in the limit of small $T$, one has
\begin{gather}
W_{\partial S} \rightarrow e^{-\frac{l^2 g^2 m S}{4}},
\end{gather}
which coincides with the result of Kogan and Kovner. In the limit of large T, one has
\begin{gather}
W_{\partial S} \rightarrow e^{-\frac{l^2 g^2 T S}{2}}
\end{gather}
and so the string tension increases linearly with temperature.
In fact, this too is to be expected. In general systems containing monopoles or vortices with some
discrete magnetic symmetry group, the area law persists at high temperature, as discussed in \cite{Korthals-Altes:2000gs}
and confirmed by lattice simulations, \emph{e.\ g.\ }\cite{Karkkainen:1993ch,Karsch:1995af}.
 Thus, the area law for the spatial Wilson loop is not a suitable confinement
criterion at finite temperature .

Remarkably, the same area law is found at finite temperature \emph{even for the non-compact theory.}
In this case the gauge-invariant Wilson loop can be written as
\begin{gather}
W_{\partial S} = \langle e^{i g \oint_{\partial S} A_i d x_i } \rangle = 
\langle  e^{ig \int_{S} B d S} \rangle.
\end{gather}
Proceeding as above and substituting for the non-compact kernels from (\ref{kers}) with $ m \rightarrow 0$ yields
\begin{gather}
W_{\partial S} \rightarrow e^{-\frac{g^2 T S}{2}}.
\end{gather}
Thus, \emph{even for the Abelian $U(1)$ theory, large spatial Wilson loops show area law behaviour at all non-zero temperatures, 
with string tension proportional to the temperature.} This result, which (to our knowledge) has
not appeared in the literature thus far, is exact since the density matrix is exact in this case.
It can be easily re-derived in the Lagrangian formalism using the propagator (\ref{lag}).
\section{Discussion}
The variational analysis of the compact $U(1)$ Hamiltonian at finite temperature has successfully reproduced
all the results derived by dimensional reduction of the Lagrangian in \cite{Agasian:1998wv}. The expected critical transition has
been identified (and at the expected temperature) and appears to belong to the expected universality class, \emph{viz.\ }BKT.
In addition, we have stressed the dependence of the critical temperature on the magnetic charge $n$ of the fundamental monopoles
in the system and shown that the area law is an inherent property of the Wilson loop at finite temperature.

We identify two possible extensions. Firstly, it would be desirable to show unequivocally that the phase transition belongs to the BKT
universality class (though we are not aware of any other phase transitions in two dimensional systems which are of infinite order).
In order to do so, one needs to show that the couplings $\tilde{z}$ and $T$ obey renormalisation group equations which 
(after a suitable co-ordinate transformation in the parameter space of relevant couplings) have the BKT form. The equation (\ref{fugfug})
generates one RG equation, but it is not clear how to derive another one in the Hamiltonian approach. Presumably, it would
suffice to calculate some gauge-invariant Green's function (and its scaling properties) in the theory using the minimal ansatz. 

Secondly, one could attempt a treatment of the full Georgi--Glashow model using the variational approach. 
It has been shown \cite{Dunne:2000vp} 
that in fact the deconfinement transition is driven by the unbinding of the $W^{\pm}$ Bosons at a temperature
below that of the monopole binding. Could this be shown using a variational method?

I would like to thank M. J. Bhaseen, I. I. Kogan and A. Kovner for their selfless assistance.

\end{document}